\documentclass[11pt]{article}

\usepackage[final]{acl}

\usepackage{times}
\usepackage{latexsym}

\usepackage[T1]{fontenc}

\usepackage[utf8]{inputenc}

\usepackage{microtype}

\usepackage{inconsolata}

\usepackage{graphicx}

\usepackage{amsmath, amssymb, amsfonts}
\usepackage{booktabs}
\usepackage{enumitem}
\usepackage{xcolor}
\usepackage{tabularx,multirow}
\usepackage{caption}
\usepackage{pifont}
\usepackage[table]{xcolor}

\title{UNISON: A Unified Sound Generation and Editing Framework via Deep LLM Fusion}

\author{Zhaoqing Li$^{\textbf{1}}$, Haoning Xu$^{\textbf{1}}$, Jingran Su$^{\textbf{2}}$, Yaofang Liu$^{\textbf{3}}$, Zhefan Rao$^{\textbf{4}}$, Huimeng Wang$^{\textbf{1}}$, \\ \textbf{Jiajun Deng$^{\textbf{1}}$, Tianzi Wang$^{\textbf{1}}$, Zengrui Jin$^{\textbf{5}}$, Rui Liu$^{\textbf{6\dag}}$, Haoxuan Che$^{\textbf{4\dag}}$\thanks{Project lead}, Xunying Liu$^{\textbf{1}}$\thanks{Corresponding authors}} \\
  $^{1}$The Chinese University of Hong Kong, $^{2}$The Hong Kong Polytechnic University \\
  $^{3}$City University of Hong Kong, $^{4}$The Hong Kong University of Science and Technology \\
  $^{5}$Tsinghua University, $^{6}$Huawei Research Hong Kong \\
  \texttt{\{zqli, xyliu\}@se.cuhk.edu.hk}}

\begin{document}
\maketitle

\begin{abstract}
We present \textbf{UNISON}, a latent diffusion framework that unifies speech generation, sound generation, and audio editing within a single model. A single model handles text-to-audio, text-to-speech, zero-shot speaker cloning, mixed speech-and-sound generation, scene-level audio editing, speech-in-scene editing, and timed temporal composition, all of which share a single set of weights.
Our architecture features two core designs: (1)~Layer-wise deep LLM fusion, which injects hidden states from uniformly sampled layers of a frozen MLLM into corresponding MM-DiT blocks via learned projections, providing depth-matched semantic conditioning that improves instruction following over single-layer baselines; and (2)~a unified multi-task architecture where task identity is encoded solely by a channel-wise mask and source audio is provided through VAE-encoded channel concatenation. Training is stabilized by an online GPU-side multi-task data synthesis pipeline with task-homogeneous batching and a two-stage curriculum.
With 621M--732M trainable parameters, UNISON achieves results competitive with or exceeding task-specialist models across evaluated domains, while being roughly $4\times$ smaller than comparable unified systems. Audio samples are available at: \url{https://lizhaoqing.github.io/UNISON-demo/}

\end{abstract}
\section{Introduction}

\begin{figure}[tbp]
    \centering
    \includegraphics[width=1.0\linewidth]{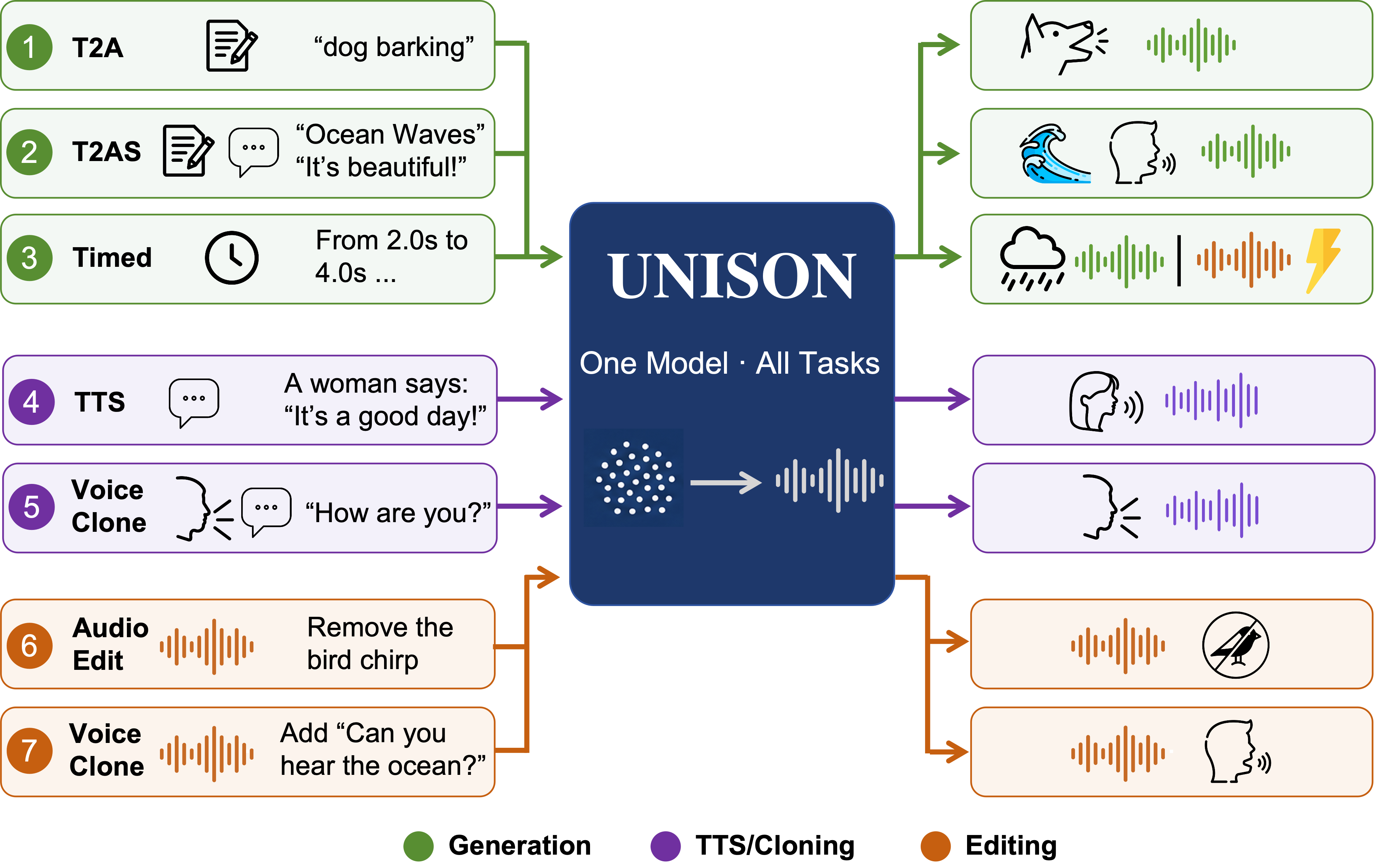}
    \vspace{-0.7cm}
    \caption{\textbf{Overview of UNISON.} A single flow-matching model handles text-to-audio generation, zero-shot TTS, gender control, audio-scene editing, and timed temporal composition. All tasks share the same architecture and weights, differentiated only by a task mask channel and optional source latent concatenation.}
    \vspace{-0.4cm}
    \label{fig:teaser}
\end{figure}

A practical audio generation system should handle diverse tasks: generating sound effects from text descriptions, synthesizing intelligible speech in a target speaker's voice, inserting or removing specific acoustic events from recordings, and composing soundscapes with temporal structure. Currently, these tasks are typically addressed by specialized models trained on isolated datasets with disparate conditioning pipelines. This fragmentation increases deployment complexity and prevents cross-task knowledge transfer---particularly between generation and editing, which differ primarily in whether a source signal is present.

Recent work has moved toward unified systems. AudioBox~\cite{audiobox} unified speech and sound generation via flow-matching with in-context masking. MMAudio~\cite{mmaudio} jointly trained on video-audio and text-audio data using an MM-DiT backbone. Audio-Omni~\cite{audioomni} expanded task coverage to include editing and multi-domain synthesis, while UniSonate~\cite{unisonate} unified TTS, music, and sound effect generation through a phoneme-driven MM-DiT. Despite these advances, two fundamental limitations persist: \textbf{(1) Inconsistent latent spaces due to task-specific auxiliary modules.} Although these systems aim for unification, most still rely on heterogeneous components for different tasks---separate mel encoders for reference audio, dedicated phoneme front-ends for TTS, separate conditioning streams for editing versus generation, or specialized duration predictors. These auxiliary modules fragment the latent space: each task operates in a subtly different representational regime, limiting cross-task knowledge transfer and complicating the training pipeline. A truly unified system should route all tasks through the \emph{same} encoder, the \emph{same} latent space, and the \emph{same} forward pass, with task identity encoded minimally.
\textbf{(2) Shallow text conditioning that discards hierarchical semantics.} A shared design choice across most existing systems is to condition the generative backbone on a single-layer text representation (typically the final hidden state of T5, CLAP, or an MLLM), which is fed identically to all DiT layers. Probing studies on transformer language models have shown that representations are organized hierarchically: lower layers primarily encode lexical and syntactic information, while higher layers capture more abstract semantic content~\cite{tenney2019,jawahar2019}. Feeding only the final-layer embedding into a generative model discards this hierarchy, potentially limiting instruction-following capacity for compositionally complex audio prompts that simultaneously specify speaker attributes, acoustic events, and temporal structure.

To address these problems, we propose \textbf{UNISON} with the following contributions:

\textbf{A unified generation-and-editing multi-task architecture with an efficient online training pipeline.} We design an architecture where all tasks (including the generation and editing of both speech and sound) share the exact same VAE, DiT backbone, and forward pass. Task identity is encoded by a single mask channel concatenated with the audio latent; source/reference audio is provided through the same frozen VAE used for targets. We build an online GPU-side data synthesis pipeline that constructs all task variations on-the-fly with task-homogeneous batching and a two-stage curriculum, enabling stable joint training of generation and editing objectives within one model.

\textbf{Layer-wise deep LLM fusion for enhanced instruction following.} We inject hidden states from uniformly sampled layers of a frozen Qwen2.5-Omni-7B text backbone into the corresponding MM-DiT double-stream blocks via learned linear projections. This provides depth-matched conditioning. Specifically, early DiT blocks receive shallow LLM representations encoding lexical and phonetic structure, while later blocks process abstract semantic features. This hierarchical alignment improves text adherence across tasks (validated in ablations, \S\ref{sec:ablation}).

\textbf{Comprehensive evaluation across diverse audio tasks.} We evaluate UNISON across multiple benchmarks spanning T2A, TTS, zero-shot cloning, mixed generation, audio editing, speech-in-scene editing, and timed composition, demonstrating that a single checkpoint achieves competitive or superior results compared to task-specialist models across all evaluated domains.

\vspace{-0.1cm}
\section{Related Work}
\vspace{-0.2cm}

\subsection{Audio and Speech Generation}
\vspace{-0.1cm}
Text-conditioned sound generation has converged on latent diffusion and flow matching~\cite{flowmatching,rectifiedflow}. AudioLDM~\cite{audioldm} and AudioGen~\cite{audiogen} pioneered text-to-audio with latent diffusion and autoregressive token modeling, respectively; AudioLDM~2~\cite{audioldm2}, Make-An-Audio~2~\cite{makeanudio2}, TangoFlux~\cite{tangoflux}, GenAU~\cite{genau}, and MMAudio~\cite{mmaudio} progressively improve quality through larger DiT~\cite{dit} models, preference optimization, or joint video-audio training, yet all condition the DiT on a \emph{single} text layer (final T5/CLAP/LLM hidden state fed identically to every block). In TTS, neural codec language models such as VALL-E~\cite{valle} demonstrated zero-shot cloning via in-context learning, inspiring modern flow-matching systems (F5-TTS~\cite{f5tts}, E2-TTS~\cite{e2tts}, MaskGCT~\cite{maskgct}, CosyVoice~\cite{cosyvoice,cosyvoice2}, ZipVoice~\cite{zipvoice}) that achieve strong quality but still rely on task-specific text front-ends such as phoneme encoders, character encoders, or duration predictors. UniSonate~\cite{unisonate} unifies TTS, T2A, and music with Qwen2.5-7B but still requires G2P phonemes and [SFX] tokens, and does not support cloning or editing.

UNISON differs in three ways: (i)~Instead of phoneme/G2P pipelines, UNISON's transcripts are plain-text LLM instructions, with zero-shot speakers encoded by the \emph{same} frozen VAE as targets; (ii)~It feeds \emph{per-block} projected LLM hidden states rather than a single-layer embedding, providing depth-matched semantic conditioning for compositional prompts. (iii) It unifies generation and editing: one checkpoint handles T2A, TTS, T2AS, zero-shot cloning, and scene editing via a task mask channel and VAE-encoded source latents, trained with an online multi-task pipeline, without separate heads or inversion stacks per task.

\vspace{-0.2cm}
\subsection{Audio Editing}
\vspace{-0.2cm}

Audio editing ranges from word-region speech tools (FluentSpeech~\cite{fluentspeech}, EdiTTS~\cite{editts}) to scene-level manipulation of mixed audio. UNISON focuses on the latter. ZETA~\cite{zeta} and SDEdit~\cite{sdedit} edit via DDPM inversion or noise--denoise schedules. MMEDIT~\cite{mmedit} trains an MM-DiT on synthetic pairs with a separate Qwen2-Audio~\cite{qwenaudio} encoder. Audio-Omni~\cite{audioomni} uses hybrid MLLM cross-attention plus a mel channel for editing.

UNISON treats editing as conditional generation: The source audio is encoded with the same VAE as a channel-concatenated input latent specified by a condition mask. This avoids inversion, auxiliary mel encoders, and task-specific decoders while preserving spectral detail in the latent domain.

\vspace{-0.2cm}
\subsection{Unified Architectures and Representation Fusion}
\vspace{-0.2cm}

UniAudio~\cite{uniaudio} and AudioBox~\cite{audiobox} unify multiple tasks via next-token prediction and flow-matching infilling, respectively. The closest concurrent systems are Audio-Omni~\cite{audioomni} and UniSonate~\cite{unisonate}. Audio-Omni (3.05B DiT) feeds only the \emph{penultimate} MLLM layer through cross-attention and routes mel/video through a second stream, which separates semantics from low-level cues but duplicates conditioning paths; its TTS is primarily evaluated on English. UniSonate (1.30B) uses last-layer Qwen features with phoneme-driven MM-DiT and omits editing and reference-based cloning. As summarized in Table~\ref{tab:arch_diff} (Appendix~\ref{app:arch_comp}), UNISON (621M--732M) is, to our knowledge, the first to jointly offer layer-wise deep fusion, plain-text bilingual zero-shot TTS, scene-level editing, and timed composition in one MM-DiT without phoneme or mel side-encoders.

On the representation side, routing frozen LLM hidden states layer-by-layer into a DiT improves text--image alignment~\cite{tangdeepfusion} and scales to large visual generators~\cite{hidream}; BAGEL~\cite{bagel} further interleaves language and visual tokens. Audio instructions are often more compositional than image captions (speaker + lexicon + background + timestamps), making depth-matched fusion particularly beneficial. UNISON is the first to apply this principle inside a unified audio model covering generation, cloning, and editing.

\vspace{-0.2cm}
\section{Method}
\vspace{-0.1cm}

\begin{figure*}[htbp]
    \centering
    \includegraphics[width=0.8\textwidth]{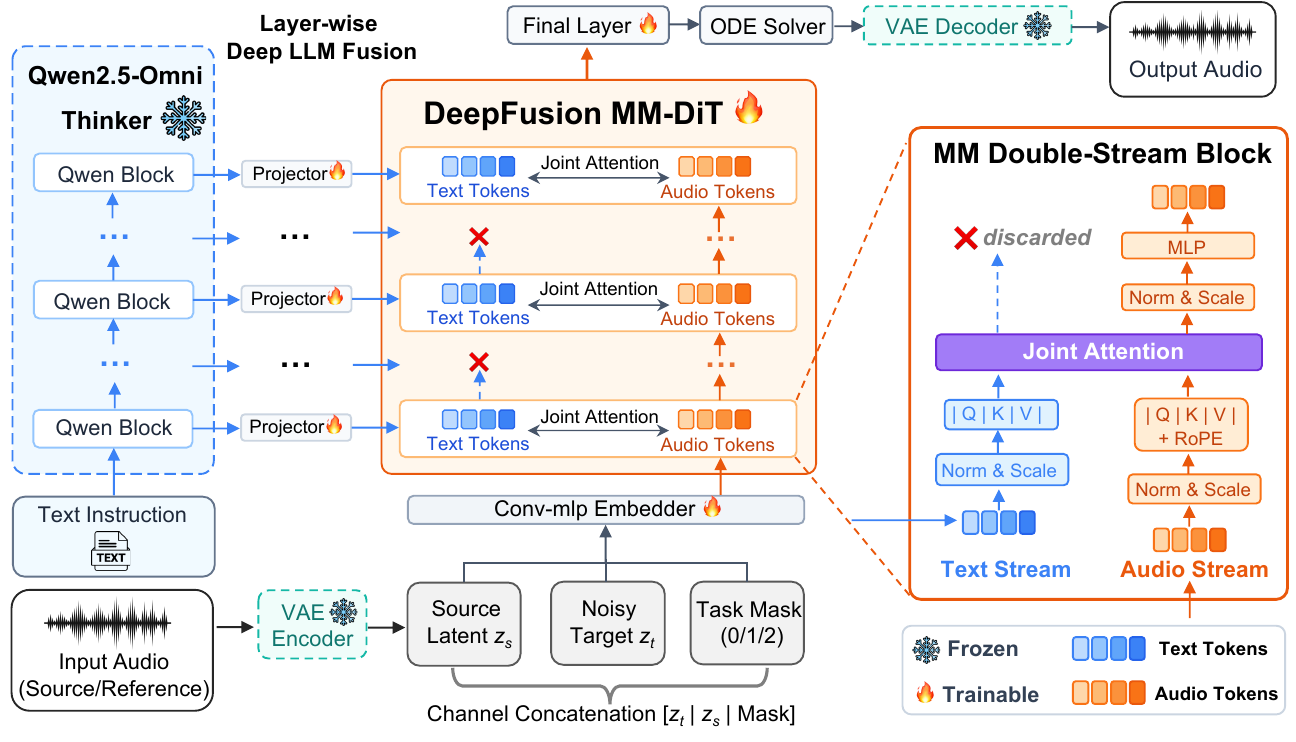}
    \vspace{-0.2cm}
    \caption{\textbf{UNISON Architecture.} \textbf{Left:} Layer-wise deep LLM fusion injects per-layer Qwen hidden states into corresponding DiT blocks via learned projectors. \textbf{Middle:} Each double-stream block performs joint attention; text tokens are refreshed per block (\ding{55}) while audio tokens pass through the MLP. \textbf{Bottom:} $[\mathbf{z}_t \,\|\, \mathbf{z}_s \,\|\, \mathbf{m}]$ are channel-concatenated and embedded; the ODE solver denoises the latent, which is VAE-decoded to waveform. See \S\ref{sec:mmdit}.}
    \label{fig:architecture}
    \vspace{-0.5cm}
\end{figure*}

Figure~\ref{fig:architecture} and the following subsections describe the architecture for generation, editing, and TTS. Frozen modules (VAE, Qwen) provide latents and text features; the trainable DeepFusion MM-DiT predicts a flow-matching velocity field.

\subsection{Overview and notation}
\label{sec:overview}

\textbf{End-to-end pipeline.} For each training sample we (i)~build a text instruction and optional source waveform, (ii)~encode waveforms with the frozen VAE to obtain $\mathbf{z}$ (target) and $\mathbf{z}_s$ (source/reference), (iii)~sample flow time $t$ and form the noisy target $\mathbf{z}_t$, (iv)~concatenate $[\mathbf{z}_t \,\|\, \mathbf{z}_s \,\|\, \mathbf{m}]$ and embed it to audio tokens $\mathbf{h}_0$, (v)~run the frozen LLM once and feed per-block text conditions $\tilde{\mathbf{E}}_k$ into the trainable MM-DiT, and (vi)~predict the velocity $v_\theta$ and backpropagate a flow-matching loss. Inference repeats step (vi) with an ODE solver starting from noise, then VAE-decodes the denoised latent.

\textbf{Symbols.} $C$ and $T'$ denote VAE latent channels and frames; $d{=}1024$ is the DiT token dimension; $D$ is the number of double-stream blocks; $L{=}28$ is the number of Qwen layers; $N$ is the instruction length in tokens.

\begin{center}
\small
\vspace{-0.3cm}
\resizebox{\linewidth}{!}{%
\begin{tabular}{llp{5.5cm}}
\toprule
\textbf{Symbol} & \textbf{Shape} & \textbf{Meaning} \\
\midrule
$\mathbf{z}$ & $C \times T'$ & Clean \textbf{target} latent (audio to generate) \\
$\mathbf{z}_s$ & $C \times T'$ & \textbf{Source} latent (zeros for pure generation) \\
$\mathbf{m}$ & $T'$ & Per-frame \textbf{task tag} in $\{0,1,2\}$ \\
$\boldsymbol{\epsilon}$ & $C \times T'$ & Gaussian noise for flow matching \\
$\mathbf{z}_t$ & $C \times T'$ & \textbf{Noised} target at flow time $t$ \\
$\mathbf{X}$ & $(2C{+}1) \times T'$ & DiT channel input $[\mathbf{z}_t \,\|\, \mathbf{z}_s \,\|\, \mathbf{m}]$ \\
$\mathbf{h}_k$ & $T' \times d$ & Audio tokens after block $k$ \\
$\mathbf{E}^{(l)}$ & $N \times 3584$ & Qwen hidden states at layer $l$ \\
$\tilde{\mathbf{E}}_k$ & $N \times d$ & Text tokens fed to DiT block $k$ \\
$v_\theta$ & $C \times T'$ & Predicted velocity (on target channels only) \\
\bottomrule
\end{tabular}
}
\vspace{-0.3cm}
\end{center}

\subsection{Audio VAE}
\label{sec:vae}

We adopt the MMAudio continuous VAE~\cite{mmaudio}. A waveform is converted to a mel spectrogram (STFT), then encoded to $\mathbf{z}$ or $\mathbf{z}_s \in \mathbb{R}^{C \times T'}$ ($C{=}40$ at 44.1\,kHz, $C{=}20$ at 16\,kHz). The same VAE encodes targets, edit sources, and speaker references, and decodes the denoised latent at inference. 
\vspace{-0.2cm}
\subsection{Multi-task inputs}
\vspace{-0.2cm}
\label{sec:task_unify}

All tasks share one network; only $(\mathbf{z}, \mathbf{z}_s, \mathbf{m})$ and the text instruction change.

\textbf{Task tag $\mathbf{m}$.} Each latent frame carries a scalar tag (broadcast as one channel in $\mathbf{X}$):
\begin{itemize}[leftmargin=*]
\vspace{-0.2cm}
    \item $\mathbf{m}{=}0$: \textit{Generation} (T2A, TTS, T2AS, timed composition). $\mathbf{z}_s{=}\mathbf{0}$.
    \vspace{-0.2cm}
    \item $\mathbf{m}{=}1$: \textit{Editing}. $\mathbf{z}_s$ is the VAE latent of the pre-edit mix; $\mathbf{z}$ is the post-edit target.
    \vspace{-0.2cm}
    \item $\mathbf{m}{=}2$: \textit{Zero-shot TTS}. $\mathbf{z}_s$ encodes the reference prefix; tags distinguish the reference region from frames to synthesize.
    \vspace{-0.2cm}
\end{itemize}

Per-task construction of $(\mathbf{z}, \mathbf{z}_s)$ and instruction templates is summarized in Table~\ref{tab:synthesis} (Appendix~\ref{app:synthesis}).

\textbf{Duration.} We use fixed-length padded latents without a separate duration head; trailing silence is learned implicitly. Timed events are specified in the text and parsed by the LLM.

\subsection{DeepFusion MM-DiT}
\label{sec:mmdit}

Building on the channel input $\mathbf{X}$ from \S\ref{sec:task_unify}, the trainable backbone is a flow-matching~\cite{flowmatching,rectifiedflow} MM-DiT~\cite{dit,mmaudio} with \textbf{layer-wise deep LLM fusion}~\cite{tangdeepfusion}. It maps $(\mathbf{X}, \{\tilde{\mathbf{E}}_k\}, t)$ to a velocity $v_\theta$ on the target channels. A default config: $D{=}20$ blocks (denoted as D20), $d{=}1024$, 8 heads.

\textbf{Noising and channel input.} Given clean $\mathbf{z}$ and noise $\boldsymbol{\epsilon}$, we sample $\sigma_t$ and set $\mathbf{z}_t = (1-\sigma_t)\mathbf{z} + \sigma_t\boldsymbol{\epsilon}$. The DiT sees
\begin{equation}
\mathbf{X} = [\mathbf{z}_t \,\|\, \mathbf{z}_s \,\|\, \mathbf{m}] \in \mathbb{R}^{(2C+1) \times T'}.
\end{equation}

A Conv-MLP embedder $\mathcal{E}$ maps $\mathbf{X}$ to $\mathbf{h}_0 \in \mathbb{R}^{T' \times d}$ (one token per frame). We \emph{zero-initialize the weights} in $\mathcal{E}$ that connect $\mathbf{z}_s$ and $\mathbf{m}$ to the token space, while $\mathbf{z}_t$ uses a standard initialization---so early training behaves like denoising-only, and the model gradually learns to use source and task channels.

\textbf{Text conditioning (left branch in Fig.~\ref{fig:architecture}).} A frozen Qwen2.5-Omni-7B~\cite{qwen25omni} Thinker runs once on the instruction, returning $\mathbf{E}^{(l)}$ for $l=1,\ldots,L$. Since the number of DiT blocks $D$ may differ from $L$, we uniformly pick $i_k = \left\lfloor 1 + k \cdot \frac{L - 1}{D - 1} \right\rfloor$ to get 
\begin{equation}
\vspace{-0.1cm}
\tilde{\mathbf{E}}_k = \mathbf{E}^{(i_k)} \mathbf{W}_k,
\vspace{0.1cm}
\end{equation}
where $\mathbf{W}_k \in \mathbb{R}^{3584 \times d}$ is the corresponding Linear projector. This ensures shallow DiT blocks see shallow Qwen layers (lexical/syntax), deep blocks see deeper semantics~\cite{tenney2019}.

\textbf{Double-stream block (right branch in Fig.~\ref{fig:architecture}).} Block $k$ receives $(\mathbf{h}_k, \tilde{\mathbf{E}}_k)$. AdaLN injects $t$; \emph{joint attention} lets all audio and text tokens attend to each other. Only $\mathbf{h}_k$ is updated by the MLP to form $\mathbf{h}_{k+1}$. We do \textbf{not} pass $\tilde{\mathbf{E}}_k$ to the next block; instead, each depth receives a fresh $\tilde{\mathbf{E}}_k$ from Qwen. Because $\tilde{\mathbf{E}}_k$ already encodes rich semantics from the frozen LLM, the DiT injects language information without relearning the full language structure. Skipping a text MLP also saves compute. RoPE~\cite{rope} applies to $\mathbf{h}_k$ only (indices $0,\ldots,T'{-}1$); QK-norm stabilizes attention.

\textbf{Output.} A linear head maps $\mathbf{h}_D$ to $v_\theta \in \mathbb{R}^{C \times T'}$ (target channels only; $\mathbf{z}_s$ and $\mathbf{m}$ are not predicted).

\subsection{Training and inference}
\label{sec:training}

\textbf{Loss.} With target velocity $\mathbf{u} = \boldsymbol{\epsilon} - \mathbf{z}$, we minimize
\begin{equation}
\mathcal{L} = \mathbb{E}_{t,\mathbf{z},\boldsymbol{\epsilon}}
\left[ \left\| v_\theta(\mathbf{X}, \{\tilde{\mathbf{E}}_k\}, t) - \mathbf{u} \right\|_2^2 \odot \mathbf{M}_{\text{loss}} \right],
\end{equation}
where $\mathbf{M}_{\text{loss}}$ zeroes the reference prefix in zero-shot TTS so gradients apply only to frames to synthesize. Text conditions are dropped with probability 0.1 for classifier-free guidance~\cite{cfg}.

\textbf{Inference.} Starting from $\mathbf{z}_t \leftarrow \boldsymbol{\epsilon}$, we integrate the learned velocity with a 100-step Euler ODE solver, then VAE-decode the denoised $\mathbf{z}$ to waveform. At inference we use CFG scale $\omega{=}4.5$.

\subsection{Online multi-task data synthesis}

Rather than constructing static datasets for each task, we implement a GPU-side online synthesis pipeline that constructs task-specific tuples on-the-fly from raw audio and speech clips. Table~\ref{tab:synthesis} in Appendix~\ref{app:synthesis} summarizes, for every task, how $\mathbf{z}_s$, $\mathbf{z}$, and the instruction template are assembled. The pipeline handles RMS normalization for SNR-controlled mixing, boundary fade-in/out, and randomized temporal offsets; instructions are assembled from predefined template pools.

\subsection{Curriculum Training with Homogeneous Batching}

To prevent gradient conflicts between generation and editing objectives, we employ:

\textbf{Two-stage curriculum.} Stage~1 trains only on generation tasks (T2A, TTS, zero-shot TTS, T2AS) for the first 150K steps, establishing a stable generative prior. Stage~2 introduces all editing tasks with the full task probability distribution (approximately 70\% generation, 30\% editing).

\textbf{Task-homogeneous batching.} Each mini-batch contains samples from a single task type, preventing intra-batch gradient conflicts between opposing objectives (e.g., ``add event'' vs. ``remove event'').

\section{Experiments}

\subsection{Implementation Details}

\textbf{Model configurations.} We train two variants: (1)~\textbf{D20} (44kHz): 20 double-stream blocks, 40 latent channels, 44.1\,kHz MMAudio VAE, 621M parameters; (2)~\textbf{D24} (16kHz): 24 double-stream blocks, 20 latent channels, 16\,kHz MMAudio VAE, 732M parameters. Both use the same frozen Qwen2.5-Omni-7B text encoder.

\textbf{Training data.} We train on a combined corpus of approximately 36M clips ($\sim$57K hours; 2.3M for audio and 33.7M for speech); per-dataset details are listed in Table~\ref{tab:train_data} (Appendix~\ref{app:train_data}). 

\textbf{Training configuration.} AdamW optimizer ($\beta_1{=}0.9$, $\beta_2{=}0.95$), learning rate $10^{-4}$ with cosine decay and 2000-step warmup, weight decay 0.01, gradient clipping 1.0. Batch size 56 per GPU on 8$\times$H800. BF16 mixed precision. EMA with decay 0.999, updated every 10 steps. CFG dropout probability 0.1. Base models trained on 10\,s max duration; fine-tuned to 22\,s for long speech only. Inference details are described in \S\ref{sec:training}.

\begin{table*}[htbp]
\centering
\footnotesize
\setlength{\tabcolsep}{13pt}
\caption{Text-to-Audio on AudioCaps~\cite{audiocaps} test set (881 clips). GT CLAP = 0.526, GT IS = 11.25. Models marked with $^\dagger$ use substantially larger training data or preference optimization.}
\vspace{-0.2cm}
\label{tab:t2a}
\resizebox{\linewidth}{!}{%
\begin{tabular}{lccccccc}
\toprule
\textbf{Model} & \textbf{Architecture} & \textbf{Params} & \textbf{FAD}$\downarrow$ & \textbf{FD}$\downarrow$ & \textbf{KL}$\downarrow$ & \textbf{IS}$\uparrow$ & \textbf{CLAP}$\uparrow$ \\
\midrule
AudioLDM 2-Large & UNet & 712M & 3.097 & 29.68 & 1.490 & 7.98 & 0.452 \\
Tango & UNet & 866M & 1.846 & 24.52 & \underline{1.305} & 7.45 & 0.498 \\
Stable Audio Open & DiT & 1.06B & 10.83 & 52.03 & 3.049 & 6.13 & 0.203 \\
Make-An-Audio 2 & DiT & 937M & 2.142 & 20.14 & 1.597 & 10.02 & 0.441 \\
GenAU-L$^\dagger$ & DiT & 1.25B & \underline{1.591} & 18.41 & \textbf{1.290} & 11.94 & \textbf{0.561} \\
\midrule
Audio-Omni & DiT & 3.05B & 2.535 & 31.42 & 1.337 & 9.55 & 0.486 \\
MMAudio-L & MM-DiT & 1.03B & 5.893 & 16.53 & 1.421 & \underline{11.98} & 0.441 \\
UniSonate & MM-DiT & 1.34B & 4.210 & 30.21 & 2.440 & 8.22 & $-$ \\
\midrule
\rowcolor{gray!15} \textbf{UNISON (D24, 16kHz)} & MM-DiT & 732M & \textbf{1.558} & \underline{16.28} & 1.459 & 10.90 & \underline{0.503} \\
\rowcolor{gray!15} \textbf{UNISON (D20, 44kHz)} & MM-DiT & 621M & 1.756 & \textbf{15.82} & 1.455 & \textbf{12.04} & 0.467 \\
\bottomrule
\end{tabular}%
}
\vspace{-0.4cm}
\end{table*}

\subsection{Evaluation Setup}
\label{sec:eval_data}

\textbf{Metrics.} We use FAD (VGGish) and FD (PANNs~\cite{panns}) for distributional quality; KL divergence and IS for classifier-based evaluation; CLAP (LAION-CLAP~\cite{laionclap}) for text--audio semantic alignment; WER/CER via Whisper-large-v3~\cite{whisper} (EN) and Paraformer~\cite{paraformer} (ZH) for intelligibility; LSD for spectral fidelity against reference targets; gender accuracy is evaluated via wav2vec2-large-XLSR-53~\cite{wav2vec2xlsr} fine-tuned on LibriSpeech for gender recognition\footnote{\url{https://huggingface.co/alefiury/wav2vec2-large-xlsr-53-gender-recognition-librispeech}}; and speech removal rate via Silero-VAD~\cite{silerovad}.

\textbf{Evaluation data.} For T2A and TTS we use standard benchmarks: AudioCaps~\cite{audiocaps} test (881 clips) and Seed-TTS~\cite{seedtts} test (1088 EN + 2020 ZH). For tasks without public benchmarks, we construct evaluation sets with a fixed seed: T2AS (600 samples, Seed-TTS speech + non-speech AudioCaps SFX at 0\,dB); audio editing (1200, 400/sub-task, non-speech AudioCaps pairs at random SNR $\in[-3,3]$\,dB); speech-in-scene editing (600, 200/sub-task, AudioCaps backgrounds + Seed-TTS speech at 10\,dB); gender TTS (300, balanced gender assignment); timed composition (150, 2--3 segment timelines). ``GT CLAP" values in tables denote the CLAP of the pseudo-GT against its caption, serving as an empirical ground truth reference since the pseudo-GT itself is artificially mixed.

\textbf{Baselines.} For T2A: AudioLDM~2~\cite{audioldm2}, Tango~\cite{tango}, Stable Audio Open~\cite{stableaudio}, Make-An-Audio~2~\cite{makeanudio2}, GenAU-L~\cite{genau}, Audio-Omni~\cite{audioomni}, MMAudio-L~\cite{mmaudio}, UniSonate~\cite{unisonate}. For TTS: MaskGCT~\cite{maskgct}, CosyVoice~2~\cite{cosyvoice2}, ZipVoice~\cite{zipvoice}, E2-TTS~\cite{e2tts}, F5-TTS~\cite{f5tts}, InstructAudio~\cite{instructaudio}, UniSonate, Audio-Omni. For editing: SDEdit~\cite{sdedit}, ZETA~\cite{zeta}, MMEDIT~\cite{mmedit}, Audio-Omni. TangoFlux~\cite{tangoflux} is excluded because its preference optimization (CRPO) is orthogonal to architecture; GenAU-L is included for reference despite its 20$\times$ larger training set.

\subsection{Main Results}

\subsubsection{Text-to-Audio Generation}

As in Table~\ref{tab:t2a}, UNISON (D24, 16\,kHz) achieves the best FAD (1.558) and CLAP (0.503) among comparable models, while D20 (44\,kHz) obtains the lowest FD (15.82) and highest IS (12.04). Both outperform Audio-Omni (3.05B) and MMAudio-L (1.03B) on FAD despite being smaller. The low FD and high CLAP scores validate the effectiveness of layer-wise deep LLM fusion for semantic alignment (further confirmed by Table~\ref{tab:ablation}, where the L-only variant shows lower CLAP). The two variants show complementary strengths: D24's larger capacity favors FAD/CLAP, while D20's 44.1\,kHz VAE better preserves spectral detail for FD/IS. GenAU-L achieves higher CLAP (0.561) but uses a 20$\times$ larger audio dataset and is a single-task model without editing or TTS capability.

\subsubsection{Text-to-Speech}
\vspace{-0.2cm}
\begin{table}[htbp]
\centering
\caption{TTS results on Seed-TTS test set. \textit{Pure TTS}: instruction-based generation without speaker reference. \textit{ZS TTS}: speaker cloning from a reference utterance.}
\label{tab:tts}
\resizebox{\linewidth}{!}{%
\begin{tabular}{llcccc}
\toprule
\multirow{2}{*}{\textbf{Model}} & \multirow{2}{*}{\textbf{Params}} & \multicolumn{2}{c}{\textbf{English}} & \multicolumn{2}{c}{\textbf{Chinese}} \\
\cmidrule(lr){3-4} \cmidrule(lr){5-6}
& & Pure WER$\downarrow$ & ZS WER$\downarrow$ & Pure CER$\downarrow$ & ZS CER$\downarrow$ \\
\midrule
MaskGCT & 1.05B & $-$ & 2.62 & $-$ & 2.27 \\
CosyVoice 2 & 618M & $-$ & 2.57 & $-$ & 1.45 \\
ZipVoice & 123M & $-$ & \underline{1.70} & $-$ & \underline{1.40} \\
E2-TTS & 333M & $-$ & 2.19 & $-$ & 1.97 \\
F5-TTS & 336M & $-$ & 1.83 & $-$ & 1.56 \\
InstructAudio & 1.30B & 1.52 & $-$ & 1.35 & $-$ \\
UniSonate & 1.34B & 1.47 & $-$ & 1.25 & $-$ \\
Audio-Omni & 3.05B & \underline{1.35} & 1.77 & $-$ & $-$ \\
\midrule
\rowcolor{gray!15} \textbf{UNISON (D24)} & 732M & \textbf{1.27} & \textbf{1.50} & \textbf{0.92} & \textbf{0.89} \\
\rowcolor{gray!15} \textbf{UNISON (D20)} & 621M & 1.42 & 1.80 & \underline{1.11} & 1.71 \\
\bottomrule
\end{tabular}%
}
\vspace{-0.4cm}
\end{table}

As shown in Table~\ref{tab:tts}, UNISON (D24) achieves the lowest error rates across all settings: pure WER 1.27\% (EN), CER 0.92\% (ZH), zero-shot WER 1.50\% and CER 0.89\%. It outperforms Audio-Omni (3.05B, pure WER 1.35\%) despite being $\sim$4$\times$ smaller, and surpasses dedicated TTS models such as ZipVoice (ZS WER 1.70\%) and F5-TTS (ZS WER 1.83\%). Notably, UNISON does not use a phoneme encoder (text conditioning is provided entirely via the frozen LLM), yet it matches models that rely on explicit G2P pipelines (UniSonate, MaskGCT). The D20 variant shows slightly higher WER (1.42\% pure, 1.80\% ZS), attributable to its smaller capacity and the increased modeling difficulty of 44.1\,kHz audio. These results confirm that multi-task training does not degrade TTS quality.

\vspace{-0.cm}
\subsubsection{Gender-Controlled TTS}
\vspace{-0.cm}
As shown in Table~\ref{tab:gender_tts}, both variants achieve perfect gender accuracy (300/300) solely from text instructions (e.g., ``A male voice saying...'') without requiring explicit speaker embeddings or gender labels during training. WER/CER remain low (D24: 1.21\% EN, 1.00\% ZH; D20: 1.47\% EN, 1.02\% ZH), indicating that gender control introduces no intelligibility degradation compared to the standard TTS setting (Table~\ref{tab:tts}).

\begin{table}[htbp]
\centering
\caption{Gender-controlled TTS on a balanced bilingual test set (300 samples: 106 EN, 194 ZH, 150 male / 150 female). Test prompts from Seed-TTS~\cite{seedtts} with randomly assigned gender. Gender accuracy evaluated via wav2vec2-large-XLSR-53 fine-tuned on LibriSpeech~\cite{wav2vec2xlsr}.}
\label{tab:gender_tts}
\resizebox{\columnwidth}{!}{%
\begin{tabular}{lcc}
\toprule
\textbf{Metric} & \textbf{UNISON (D24, 16kHz)} & \textbf{UNISON (D20, 44kHz)} \\
\midrule
Gender Accuracy$\uparrow$ & 100\%  & 100\% \\
WER-EN$\downarrow$ & 1.21 & 1.47 \\
CER-ZH$\downarrow$ & 1.00 & 1.02 \\
WER (male)$\downarrow$ & 1.64 & 1.31 \\
WER (female)$\downarrow$ & 0.74 & 1.65 \\
CER (male)$\downarrow$ & 1.30 & 1.05 \\
CER (female)$\downarrow$ & 0.71 & 0.99 \\
\bottomrule
\end{tabular}
}
\end{table}

\vspace{-0.2cm}
\subsubsection{Mixed Speech + Sound Generation}

As shown in Table~\ref{tab:t2as}, UNISON (D24) achieves CLAP 0.444 (93.3\% of the pseudo-GT CLAP of 0.476), with WER 2.04\% and CER 3.64\% measured directly on the mixed output without source separation. The D20 variant shows slightly lower speech clarity (WER 3.44\%, CER 5.80\%) but achieves lower LSD (2.36 vs.\ 2.44), indicating better waveform-level fidelity to the pseudo-GT mixture. This task is unique in that no existing public benchmark or baseline exists for single-model joint speech+sound generation; UNISON handles it naturally by leveraging multi-task training on both T2A and TTS data without any dedicated mixing module or two-stage pipeline.

\begin{table}[htbp]
\centering
\caption{T2AS: generating a unified output containing intelligible speech and a matching background soundscape from a joint instruction. The test set (600 samples) is constructed by pairing Seed-TTS speech entries with AudioCaps sound clips, mixed at 0\,dB SNR as pseudo ground-truth. GT CLAP is computed on this pseudo-GT against the evaluation caption. WER/CER are measured directly on the mixed output (not separated). LSD is computed against the pseudo-GT.}
\label{tab:t2as}
\resizebox{\columnwidth}{!}{%
\begin{tabular}{lcc}
\toprule
\textbf{Metric} & \textbf{UNISON (D24, 16kHz)} & \textbf{UNISON (D20, 44kHz)} \\
\midrule
CLAP$\uparrow$ (GT: 0.476) & 0.444 & 0.430 \\
WER-EN$\downarrow$ & 2.04 & 3.44 \\
CER-ZH$\downarrow$ & 3.64 & 5.80 \\
LSD$\downarrow$ & 2.44 & 2.36 \\
\bottomrule
\end{tabular}%
}
\vspace{-0.5cm}
\end{table}

\subsubsection{Audio Editing}

As shown in Table~\ref{tab:audio_edit}, UNISON (D24) achieves the best FD and CLAP across all sub-tasks, with overall FD 12.38 (vs.\ 20.60 for MMEDIT) and CLAP 0.364 (vs.\ 0.257), reaching 82\% of the pseudo-GT CLAP. LSD remains $\leq$2.15 across all sub-tasks, confirming preservation of non-edited content. The ``Remove'' sub-task shows lower CLAP for all methods (D24: 0.308, MMEDIT: 0.221), reflecting the difficulty of spectral disentanglement. D20 achieves lower LSD due to its higher-bandwidth VAE but shows higher FD and lower CLAP.

By encoding source audio through the same frozen VAE used for the target, UNISON operates in a shared latent space, unlike SDEdit/ZETA (noise-injection/inversion) or Audio-Omni (separate mel encoder). Qualitative mel spectrograms are provided in Appendix~\ref{app:edit_qual} (Figs~\ref{fig:edit_audio_mel_d24}--\ref{fig:edit_audio_mel_d20}).

\begin{table}[t!]
\centering
\caption{Audio editing on 1200 constructed test samples (400 per sub-task). Source/target pairs are synthesized by mixing AudioCaps~\cite{audiocaps} test clips at random SNR (see \S\ref{sec:eval_data}). GT CLAP (in parentheses) is the CLAP score of the constructed target against the evaluation caption, serving as a pseudo-GT reference. LSD is computed between generated audio and the constructed target.}
\label{tab:audio_edit}
\resizebox{\linewidth}{!}{%
\begin{tabular}{llccc}
\toprule
\textbf{Task (GT CLAP)} & \textbf{Model} & \textbf{FD}$\downarrow$ & \textbf{LSD}$\downarrow$ & \textbf{CLAP}$\uparrow$ \\
\midrule
\multirow{6}{*}{Add (0.429)}
& SDEdit & 78.86 & 2.21 & 0.168 \\
& ZETA & 67.27 & 2.18 & 0.243 \\
& MMEDIT & 25.98 & 2.23 & 0.339 \\
& Audio-Omni & 34.92 & 1.99 & 0.332 \\
\rowcolor{gray!15} \cellcolor{white} & UNISON (D24) & \textbf{19.26} & \underline{1.49} & \textbf{0.416} \\
\rowcolor{gray!15} \cellcolor{white} & UNISON (D20) & \underline{20.18} & \textbf{1.43} & \underline{0.391} \\
\midrule
\multirow{6}{*}{Remove (0.485)}
& SDEdit & 87.65 & 2.11 & 0.053 \\
& ZETA & 66.34 & 2.09 & 0.141 \\
& MMEDIT & 45.25 & 3.86 & \underline{0.221} \\
& Audio-Omni & 64.00 & 2.51 & 0.112 \\
\rowcolor{gray!15} \cellcolor{white} & UNISON (D24) & \textbf{33.20} & 2.15 & \textbf{0.308} \\
\rowcolor{gray!15} \cellcolor{white} & UNISON (D20) & \underline{37.93} & 2.18 & 0.169 \\
\midrule
\multirow{6}{*}{Replace (0.417)}
& SDEdit & 79.09 & 1.90 & 0.119 \\
& ZETA & 62.71 & 1.89 & 0.180 \\
& MMEDIT & 27.56 & 2.77 & 0.210 \\
& Audio-Omni & 55.39 & 1.82 & 0.202 \\
\rowcolor{gray!15} \cellcolor{white} & UNISON (D24) & \textbf{21.31} & \underline{1.68} & \textbf{0.368} \\
\rowcolor{gray!15} \cellcolor{white} & UNISON (D20) & \underline{23.09} & \textbf{1.57} & \underline{0.307} \\
\midrule
\multirow{6}{*}{Overall (0.444)}
& SDEdit & 73.85 & 2.07 & 0.114 \\
& ZETA & 57.27 & 2.05 & 0.189 \\
& MMEDIT & 20.60 & 2.95 & 0.257 \\
& Audio-Omni & 36.29 & 2.11 & 0.217 \\
\rowcolor{gray!15} \cellcolor{white} & UNISON (D24) & \textbf{12.38} & \underline{1.77} & \textbf{0.364} \\
\rowcolor{gray!15} \cellcolor{white} & UNISON (D20) & \underline{13.44} & \textbf{1.73} & \underline{0.289} \\
\bottomrule
\end{tabular}%
}
\vspace{-0.5cm}
\end{table}

\subsubsection{Speech-in-Scene Editing}

Speech-in-scene editing manipulates spoken content within an existing audio scene (speech mixed with background sounds) by inserting, deleting, or rewriting speech while preserving the non-speech background intact. Results are shown in Table~\ref{tab:speech_edit}.

D24 achieves 99.16\% speech removal (Delete) with LSD 1.56, and maintains WER $\leq$1.35\% for Insert/Rewrite, confirming effective voice suppression and high synthesized-speech intelligibility. D20 shows lower LSD across sub-tasks but lower CLAP and removal rate, consistent with the D24/D20 trade-off observed in audio editing. Qualitative examples are in Appendix~\ref{app:edit_qual} (Figs~\ref{fig:edit_speech_mel_d24}--\ref{fig:edit_speech_mel_d20}).

\begin{table}[htbp]
\centering
\small

\caption{Speech-in-scene editing (200 samples per sub-task). Test pairs constructed from AudioCaps backgrounds + Seed-TTS~\cite{seedtts} speech mixed at 10\,dB SNR. CLAP computed on pseudo-GT. WER/CER measured directly on full output. LSD computed against constructed target. Removal rate via Silero-VAD~\cite{silerovad}.}
\vspace{-0.2cm}
\label{tab:speech_edit}
\resizebox{\columnwidth}{!}{%
\begin{tabular}{lccccccc}
\toprule
\textbf{Sub-task} & \textbf{Model} & \textbf{CLAP}$\uparrow$ & \textbf{GT CLAP} & \textbf{LSD}$\downarrow$ & \textbf{WER}$\downarrow$ & \textbf{CER}$\downarrow$ & \textbf{Removal}$\uparrow$ \\
\midrule
\multirow{2}{*}{Insert} & D24 & 0.433 & 0.459 & 1.70 & 1.35 & 0.65 & $-$ \\
& D20 & 0.429 & 0.459 & 1.66 & 1.70 & 1.08 & $-$ \\
\midrule
\multirow{2}{*}{Delete} 
& D24 & 0.412 & 0.468 & 1.56 & $-$ & $-$ & 99.16\% \\
& D20 & 0.320 & 0.468 & 1.52 & $-$ & $-$ & 95.72\% \\
\midrule
\multirow{2}{*}{Rewrite} 
& D24 & 0.408 & 0.456 & 1.60 & 0.98 & 0.95 & $-$ \\
& D20 & 0.396 & 0.456 & 1.46 & 1.35 & 1.09 & $-$ \\
\bottomrule
\end{tabular}
}
\vspace{-0.4cm}
\end{table}

\subsubsection{Timed Audio Generation}
As in Table~\ref{tab:timed}, both variants achieve per-segment CLAP $\geq$0.308, with overall CLAP exceeding the per-segment value (D24: 0.345; D20: 0.405), indicating coherent holistic scenes despite some boundary softening. Temporal control relies purely on natural-language timestamp parsing by the frozen LLM without dedicated alignment modules. Mel spectrograms in Appendix~\ref{app:timed_qual} (Figs~\ref{fig:timed_mel_d24}--\ref{fig:timed_mel_d20}) confirm spectral alignment with specified time intervals.

\begin{table}[htbp]
\centering
\caption{Timed composition (150 test samples with 2--3 segment temporal instructions). Per-segment CLAP measures semantic alignment within each time window; overall CLAP measures holistic scene quality.}
\label{tab:timed}
\resizebox{\columnwidth}{!}{%
\begin{tabular}{lcc}
\toprule
\textbf{Metric} & \textbf{UNISON (D24, 16kHz)} & \textbf{UNISON (D20, 44kHz)} \\
\midrule
Per-segment CLAP$\uparrow$ & 0.308 & 0.311 \\
Overall CLAP$\uparrow$ & 0.345 & 0.405 \\
\bottomrule
\end{tabular}%
}
\vspace{-0.3cm}
\end{table}

\subsection{Ablation Studies}
\label{sec:ablation}

We conduct ablations on three axes (LLM conditioning mode, stream architecture, and LLM scale) using the same training data and hyperparameters.

\textbf{For LLM conditioning mode}, we compare three strategies on the double-stream D24 architecture: (1)~D24-O~(deep fusion only): per-block projections from uniformly sampled LLM layers, no persistent text stream, text MLP disabled; (2)~D24-L~(penultimate layer only): a single projection from the penultimate LLM layer broadcast to all DiT blocks, text MLP enabled; (3)~D24-OL~(deep + penultimate): both mechanisms active simultaneously.
\textbf{For stream architecture}, we compare D24-O (double-stream, 24 blocks, separate text/audio normalization and QKV) against S32-O (single-stream, 32 blocks, text and audio tokens share normalization, QKV projections, and MLP) with comparable FLOPs.
\textbf{For LLM scale}, We test D24-O-3B, which replaces the default 7B Qwen2.5-Omni Thinker with a 3B variant to assess the effect of LLM capacity on conditioning quality.

Table~\ref{tab:ablation} reveals several findings:

\textbf{Deep fusion improves semantic following.} D24-L (penultimate layer only) achieves the lowest CLAP (0.175) and highest FD (22.71) among D24 variants, confirming that broadcasting a single LLM layer provides weaker conditioning than depth-matched injection. Both D24-O and D24-OL achieve lower FD (20.46, 20.18) and higher CLAP (0.180, 0.187), demonstrating that per-block deep fusion better captures hierarchical text semantics for audio generation.

\textbf{Redundant text tokens hurt TTS.} D24-OL achieves the best FD (20.18) and CLAP (0.187), yet the highest WER (5.52\%). In this variant, text tokens enter the DiT from \emph{two} sources (the persistent last-layer stream and the per-block deep fusion projections), effectively duplicating the conditioning signal. This redundancy improves T2A semantic alignment but introduces noise that increases TTS difficulty. D24-O avoids this trade-off by using only per-block projections with \emph{ephemeral} text tokens, achieving competitive FD/CLAP (20.46/0.180) while maintaining the lowest WER (4.33\%).

\textbf{Double-stream architecture is essential.} S32-O (single-stream) shows the worst FD (23.19), lowest CLAP (0.169), and high WER (4.84\%), despite using the same deep fusion as D24-O. Sharing normalization and QKV projections between text and audio tokens prevents modality-specific representations; the double-stream design avoids this by maintaining separate feature spaces with interaction only through joint attention.

\textbf{LLM scale matters.} D24-O-3B shows degraded performance on all metrics (FD: 20.46$\to$21.53, CLAP: 0.180$\to$0.174, WER: 4.33$\to$5.61), confirming that richer LLM representations directly benefit both semantic following and speech intelligibility.

\begin{table}[htbp]
\centering
\caption{Ablation on AudioCaps (T2A) and Seed-TTS EN (pure TTS). All variants use the same training data and hyperparameters (80K steps). ``3B'' denotes the Qwen2.5-Omni-3B Thinker in place of the default 7B.}
\label{tab:ablation}
\resizebox{\columnwidth}{!}{%
\begin{tabular}{llcccc}
\toprule
\textbf{Conditioning} & \textbf{Arch} & \textbf{Params} & \textbf{FD}$\downarrow$ & \textbf{CLAP}$\uparrow$ & \textbf{WER-EN}$\downarrow$ \\
\midrule
\rowcolor{gray!15} O (deep fusion, 7B) & D24 & 732M & \underline{20.46} & \underline{0.180} & \textbf{4.33} \\
L (penultimate-layer, 7B) & D24 & 975M & 22.71 & 0.175 & \underline{4.44} \\
OL (deep + penultimate, 7B) & D24 & 1,063M & \textbf{20.18} & \textbf{0.187} & 5.52 \\
\midrule
O (deep fusion, 7B) & S32 & 685M & 23.19 & 0.169 & 4.84 \\
\midrule
O (deep fusion, 3B) & D24 & 694M & 21.53 & 0.174 & 5.61 \\
\bottomrule
\end{tabular}%
}
\vspace{-0.3cm}
\end{table}

\section{Conclusion}

We presented UNISON, a framework that unifies audio generation and editing through layer-wise deep LLM fusion and a channel-concatenation architecture that routes all tasks through a single VAE, DiT backbone, and forward pass. A single 621M--732M parameter checkpoint achieves competitive or superior results across T2A, TTS, zero-shot cloning, audio editing, and temporal composition without task-specific modules, demonstrating that multi-task audio generation at scale does not necessarily require heterogeneous conditioning paths. These results suggest a practical path toward general-purpose audio systems that grow in capability through data and model scaling rather than architectural specialization.

\section*{Limitations}

\textbf{VAE reconstruction quality.} UNISON relies on the pre-trained MMAudio VAE, which was originally designed for environmental sound synthesis. While it provides a compact and effective latent space for general audio, its reconstruction fidelity for speech---particularly high-frequency formant details, subtle prosodic variations, and breathy or whispered voice qualities---imposes an upper bound on overall output quality. This is especially noticeable for zero-shot TTS, where fine-grained speaker timbre nuances may be smoothed out during VAE encoding. A natural next step is to train a unified VAE with improved speech reconstruction, potentially adopting higher latent resolution or a multi-scale architecture that better preserves both spectral detail and temporal dynamics.

\textbf{Synthetic training data for editing.} Our editing and T2AS training data is constructed by algorithmically mixing open-source audio clips (RMS-based overlay with random temporal placement and fade-in/out). While this approach validates the architectural design and enables large-scale training without manual annotation, the resulting data distribution differs from naturalistic recordings in several ways: (i)~real-world audio scenes exhibit complex acoustic interactions (e.g., reverberation, occlusion, Lombard effects) that simple mixing cannot capture; (ii)~caption quality for AudioSet/WavCaps sources has not undergone rigorous human verification, introducing label noise; (iii)~the SNR distribution and temporal alignment of synthetic mixtures may not reflect typical editing scenarios encountered in practice. Future work will explore more realistic synthesis pipelines (e.g., room impulse response convolution, physically-informed mixing) and incorporate human-verified editing pairs.

\textbf{Scale and modality coverage.} The current model (621M--732M DiT parameters) is trained on $\sim$36M clips ($\sim$57K hours). Both model size and data quantity are moderate relative to recent scaling efforts such as GenAU (1.25B params, 47M clips with synthetic captions). The architecture is designed to scale: the channel-concatenation mechanism naturally extends to additional modalities (e.g., video features for V2A generation) and the deep fusion framework can accommodate larger LLM backbones. We have not yet explored these directions but anticipate substantial gains from increased scale.

\textbf{Language and domain scope.} The current system supports English and Chinese speech; extension to other languages requires additional multilingual speech data but no architectural changes. We do not target music generation in this work, primarily because large-scale, openly licensed music datasets with high-quality text annotations remain difficult to obtain due to copyright restrictions. Additionally, music generation involves distinct challenges---long-range harmonic structure, multi-instrument arrangement, and beat/tempo consistency~\cite{musicgen}---that may benefit from domain-specific design choices (e.g., hierarchical latent representations or music-aware tokenization) beyond our current scope. Nevertheless, the architecture itself is domain-agnostic and could incorporate music data if suitable training corpora become available.


\bibliography{main}
\appendix

\section{Architectural Comparison}
\label{app:arch_comp}

Table~\ref{tab:arch_diff} provides a detailed side-by-side comparison of UNISON with the two closest concurrent unified audio systems (Audio-Omni and UniSonate), highlighting differences in LLM conditioning strategy, transcript encoding, reference audio handling, task coverage, and model scale.

\begin{table*}[htbp]
\centering
\footnotesize
\caption{Architectural comparison with Audio-Omni and UniSonate.}
\label{tab:arch_diff}
\resizebox{\linewidth}{!}{%
\begin{tabularx}{\linewidth}{lXXX}
\toprule
\textbf{Aspect} & \textbf{Audio-Omni} & \textbf{UniSonate} & \textbf{UNISON (ours)} \\
\midrule
LLM conditioning & Penultimate MLLM layer $\rightarrow$ cross-attn & Qwen2.5-7B last layer $\rightarrow$ double-stream & Layer-wise: LLM layer $i$ $\rightarrow$ DiT block $i$ \\
Transcript encoding & ConvNeXt V2 character encoder & G2P phoneme + [SFX] tokens & Plain text via LLM; no extra module \\
Reference audio & Mel Encoder + channel-cat & Not supported & Same frozen VAE + channel-cat \\
Zero-shot TTS & English only (voice conversion) & Not supported & Bilingual (EN/ZH), VAE channel-cat \\
Audio editing & Hybrid stream (AudioEdit data) & Not supported & Same forward pass, mask=1 \\
Music domain & Supported & Supported & Not targeted \\
Parameters & 3.05B & 1.34B & 621M / 732M \\
\bottomrule
\end{tabularx}
}
\end{table*}

\section{Online Multi-task Data Synthesis}
\label{app:synthesis}

Table~\ref{tab:synthesis} summarizes how each task's training tuple $(\mathbf{z}_s, \mathbf{z}, \text{instruction})$ is constructed on-the-fly from raw audio and speech clips during training. All synthesis is performed on GPU at data-loading time, requiring no pre-computed static datasets.

\begin{table*}[htbp]
\centering
\caption{Online data synthesis: each task is constructed from base audio/speech clips at training time.}
\label{tab:synthesis}
\resizebox{\linewidth}{!}{
\begin{tabular}{llll}
\toprule
\textbf{Task} & \textbf{Source latent} $\mathbf{z}_s$ & \textbf{Target latent} $\mathbf{z}$ & \textbf{Instruction format} \\
\midrule
T2A (generation) & Zeros & Audio clip & \texttt{[Audio] \{caption\}} \\
TTS (gender-controlled) & Zeros & Speech clip & \texttt{[Speech] A \{gender\} voice saying "\{text\}"} \\
Zero-shot TTS & VAE(ref + zeros pad) & Full utterance & \texttt{[Speech with voice] \{text\}} \\
T2AS (mixed speech+sound) & Zeros & RMS-mixed speech + SFX & \texttt{[Speech] ... [Audio] ...} \\
Audio edit --- add & VAE(base) & Base + added event & \texttt{[Edit][Audio] Add \{event\}} \\
Audio edit --- remove & VAE(base + event) & Base only & \texttt{[Edit][Audio] Remove \{event\}} \\
Audio edit --- replace & VAE(base + old) & Base + new event & \texttt{[Edit][Audio] Remove \{old\} [Edit][Audio] add \{new\}} \\
Speech-in-scene --- insert & VAE(background) & Background + speech & \texttt{[Edit][Speech] Add a voice saying "\{text\}"} \\
Speech-in-scene --- delete & VAE(bg + speech) & Background only & \texttt{[Edit][Speech] Remove the speech} \\
Speech-in-scene --- rewrite & VAE(bg + old speech) & Background + new speech & \texttt{[Edit][Speech] Change speech to "\{new\}"} \\
Timed composition & Zeros & Multi-event timeline & \texttt{[Audio] From \{t1\}s to \{t2\}s, \{event\}...} \\
\bottomrule
\end{tabular}}
\end{table*}

\section{Training Data Composition}
\label{app:train_data}

Table~\ref{tab:train_data} lists all datasets used for training, broken down by domain (audio vs.\ speech). The combined corpus contains approximately 36M clips totaling $\sim$57K hours. Audio clips with speech-heavy captions are filtered out at load time ($\sim$390K removed); speech clips shorter than 3\,s are excluded from zero-shot TTS sampling ($\sim$29M eligible of 33.7M total).

\begin{table}[t!]
\centering
\caption{Training data composition. Sources include WavCaps~\cite{wavcaps}, AudioSet~\cite{audioset}, VGGSound~\cite{vggsound}, LibriTTS~\cite{libritts}, WenetSpeech~\cite{wenet}, and Emilia~\cite{emilia}.}
\label{tab:train_data}
\resizebox{\columnwidth}{!}{%
\begin{tabular}{llrr}
\toprule
\textbf{Domain} & \textbf{Dataset} & \textbf{Clips} & \textbf{Hours} \\
\midrule
\multirow{3}{*}{Audio} & WavCaps & 841K & 2,189 \\
& AudioSet (with labels) & 1,718K & 4,772 \\
& VGGSound & 174K & 482 \\
\cmidrule(lr){2-4}
& \textit{Audio subtotal (after filter)} & \textit{2,342K} & \textit{$\sim$6,400} \\
\midrule
\multirow{4}{*}{Speech} & LibriTTS & 281K & 332 \\
& WenetSpeech & 8,133K & 7,693 \\
& Emilia-EN (20M subset) & 11,560K & 19,802 \\
& Emilia-ZH (20M subset) & 13,759K & 22,892 \\
\cmidrule(lr){2-4}
& \textit{Speech subtotal} & \textit{33,733K} & \textit{$\sim$50,700} \\
\midrule
\multicolumn{2}{l}{\textbf{Total}} & \textbf{$\sim$36M} & \textbf{$\sim$57K} \\
\bottomrule
\end{tabular}%
}
\end{table}

\section{Task Probability Distribution}
\label{app:task_probs}

Table~\ref{tab:task_probs} shows the task sampling probabilities used during training. Stage~1 (first 150K steps) trains only on generation tasks; Stage~2 introduces editing tasks with the full probability distribution shown below.

\begin{table}[t]
\centering
\caption{Task sampling probabilities in Stage~2 (joint training).}
\resizebox{\columnwidth}{!}{%
\label{tab:task_probs}
\begin{tabular}{lcc}
\toprule
\textbf{Task} & \textbf{Stage 1} & \textbf{Stage 2 Prob.} \\
\midrule
TTS (gender-controlled) & \checkmark & 0.15 \\
Zero-shot TTS & \checkmark & 0.25 \\
T2A (single event) & \checkmark & 0.10 \\
T2A (mix) & \checkmark & 0.08 \\
Speech + audio mix (T2AS) & \checkmark & 0.08 \\
Timed composition & \checkmark & 0.04 \\
\midrule
Audio edit --- add & $\times$ & 0.04 \\
Audio edit --- remove & $\times$ & 0.04 \\
Audio edit --- replace & $\times$ & 0.03 \\
Speech-in-scene --- insert & $\times$ & 0.04 \\
Speech-in-scene --- delete & $\times$ & 0.03 \\
Speech-in-scene --- rewrite & $\times$ & 0.04 \\
Denoise & $\times$ & 0.04 \\
Other editing variants & $\times$ & 0.04 \\
\bottomrule
\end{tabular}}
\end{table}

\section{Architecture Details}
\label{app:arch}

Table~\ref{tab:arch_details} provides the complete set of architecture and training hyperparameters for both model variants (D20S0-44kHz and D24S0-16kHz), including VAE configuration, DiT dimensions, optimizer settings, and compute resources.

\begin{table*}[htbp]
\centering
\caption{Full architecture and training hyperparameters.}
\label{tab:arch_details}
\resizebox{\linewidth}{!}{
\begin{tabular}{lll}
\toprule
\textbf{Hyperparameter} & \textbf{D20S0-44kHz (primary)} & \textbf{D24S0-16kHz} \\
\midrule
Audio VAE & MMAudio 44.1kHz VAE & MMAudio 16kHz VAE \\
VAE latent channels ($C$) & 40 & 20 \\
DiT input channels (target + source + mask) & 81 (40+40+1) & 41 (20+20+1) \\
Sample rate & 44,100 Hz & 16,000 Hz \\
Max training duration & 10\,s (base) / 22\,s (fine-tune) & 10\,s (base) / 22\,s (fine-tune) \\
DiT hidden size & 1,024 & 1,024 \\
Attention heads & 8 & 8 \\
Head dimension & 128 & 128 \\
MLP width ratio & 4.0 & 4.0 \\
Double-stream blocks & 20 & 24 \\
Single-stream blocks & 0 & 0 \\
Patch embedder & \texttt{conv\_mlp} (kernel=7) & \texttt{conv\_mlp} (kernel=7) \\
RoPE dimensions & [96, 16, 16] & [96, 16, 16] \\
LLM encoder & Qwen2.5-Omni-7B (frozen) & Qwen2.5-Omni-7B (frozen) \\
LLM hidden dim & 3,584 & 3,584 \\
LLM total layers & 28 & 28 \\
Fusion mode & O (deep only) & O (deep only) \\
Layer selection & Uniform interval (28 $\rightarrow$ 20) & Uniform interval (28 $\rightarrow$ 24) \\
Per-block projector & Linear ($3584 \to 1024$) & Linear ($3584 \to 1024$) \\
Trainable parameters & 621M & 732M \\
Flow scheduler & FlowMatchEulerDiscrete & FlowMatchEulerDiscrete \\
Inference steps & 100 & 100 \\
CFG scale & 4.5 & 4.5 \\
Optimizer & AdamW ($\beta_1$=0.9, $\beta_2$=0.95) & AdamW ($\beta_1$=0.9, $\beta_2$=0.95) \\
Learning rate & 1e-4 (cosine, 2000 warmup) & 1e-4 (cosine, 2000 warmup) \\
Weight decay & 0.01 & 0.01 \\
Gradient clipping & 1.0 & 1.0 \\
Mixed precision & BF16 & BF16 \\
EMA decay & 0.999 (update every 10 steps) & 0.999 (update every 10 steps) \\
Training GPUs & 8 $\times$ H800 & 8 $\times$ H800 \\
Batch size per GPU & 56 & 56 \\
\bottomrule
\end{tabular}}
\end{table*}












\section{Editing Qualitative Examples}
\label{app:edit_qual}

\begin{figure*}[b!]
    \centering
    \includegraphics[width=\linewidth]{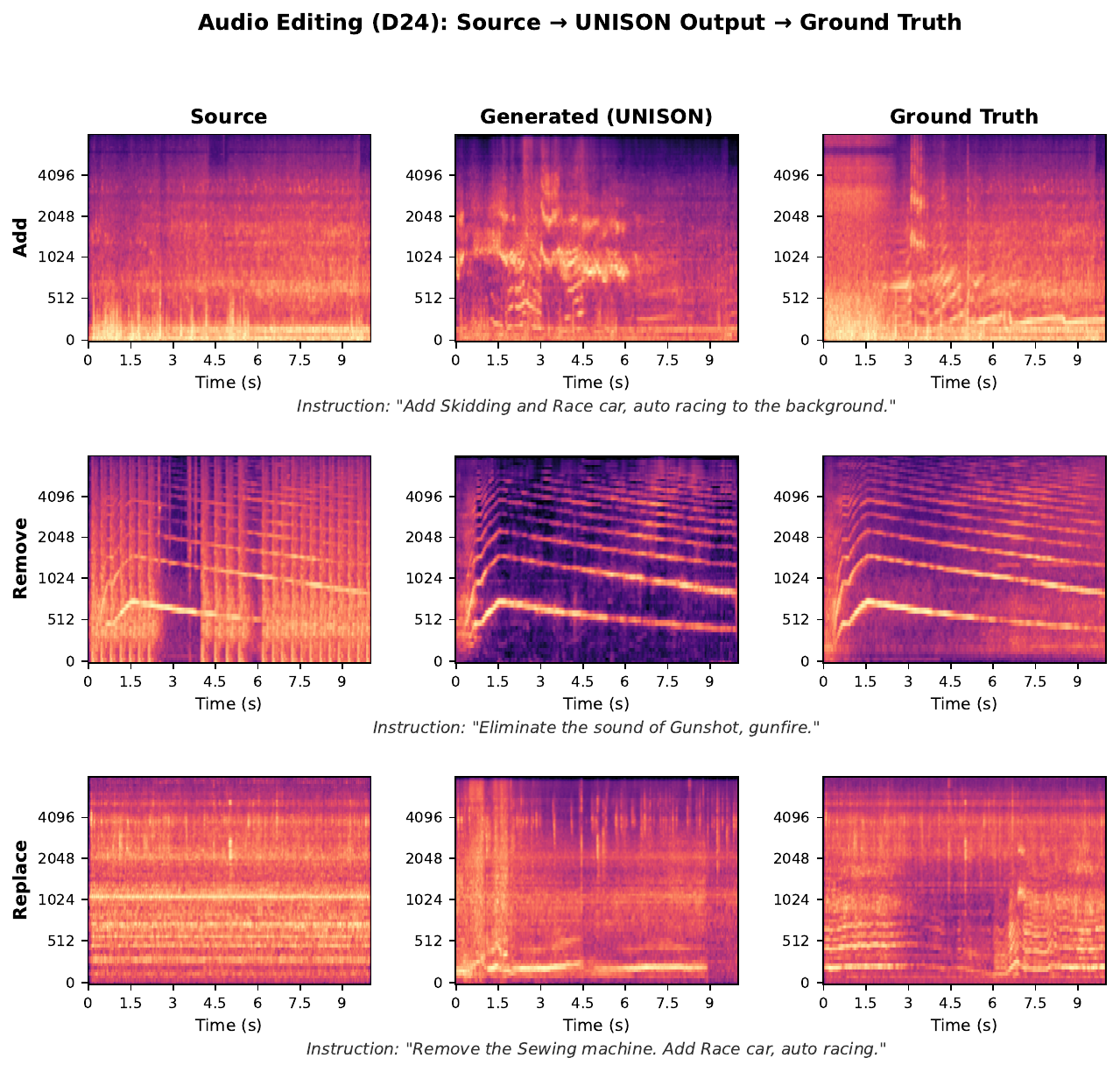}
    \caption{Audio editing qualitative examples from \textbf{UNISON (D24, 16\,kHz)}. Each row shows one sub-task (Add / Remove / Replace). Left: source audio. Middle: UNISON output. Right: constructed ground truth.}
    \label{fig:edit_audio_mel_d24}
\end{figure*}

\begin{figure*}[bp]
    \centering
    \includegraphics[width=\linewidth]{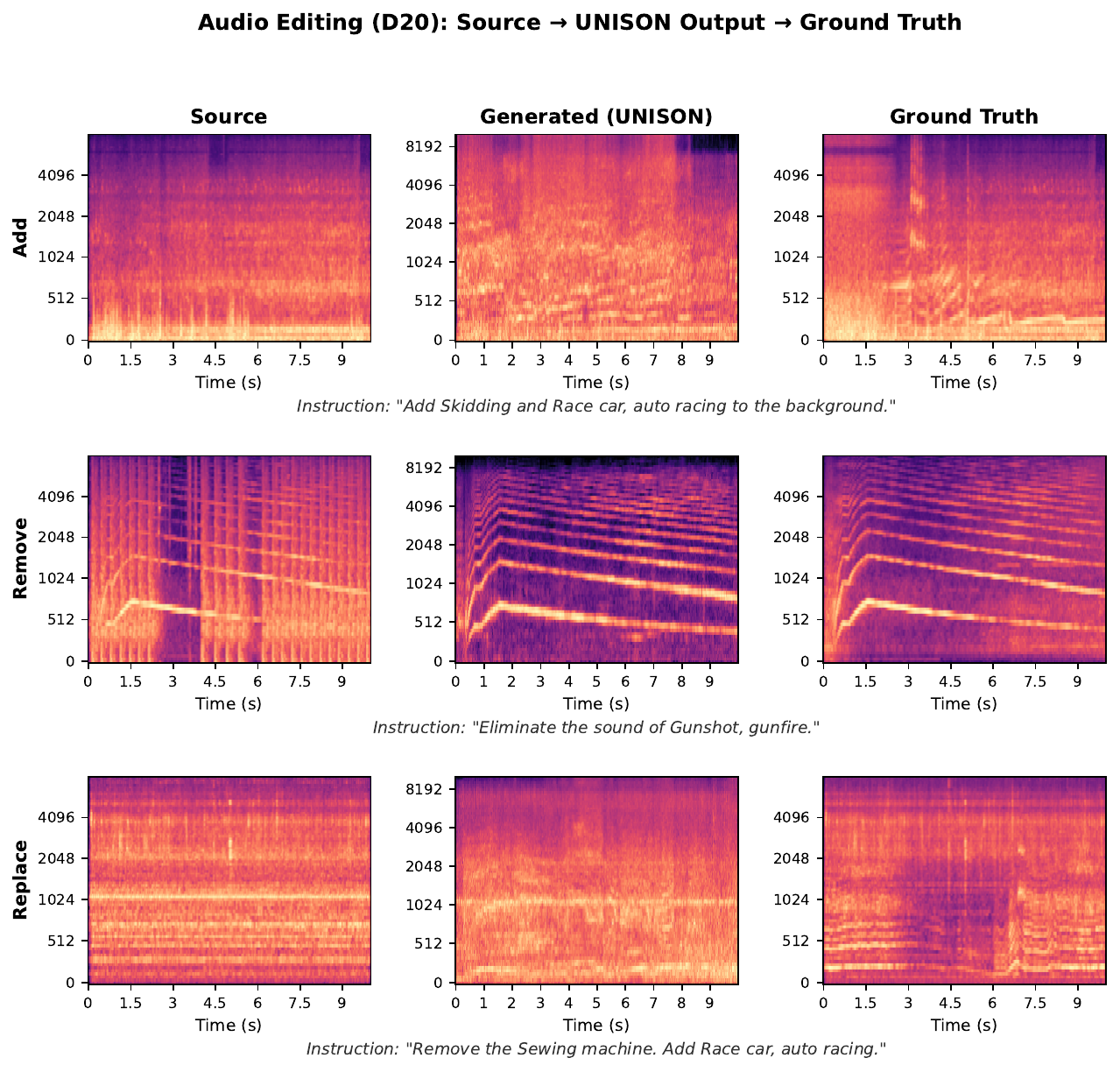}
    \caption{Audio editing qualitative examples from \textbf{UNISON (D20, 44.1\,kHz)} on the same samples as Figure~\ref{fig:edit_audio_mel_d24}. The higher-bandwidth VAE preserves more spectral detail in both source and generated outputs.}
    \label{fig:edit_audio_mel_d20}
\end{figure*}

\begin{figure*}[bp]
    \centering
    \includegraphics[width=\linewidth]{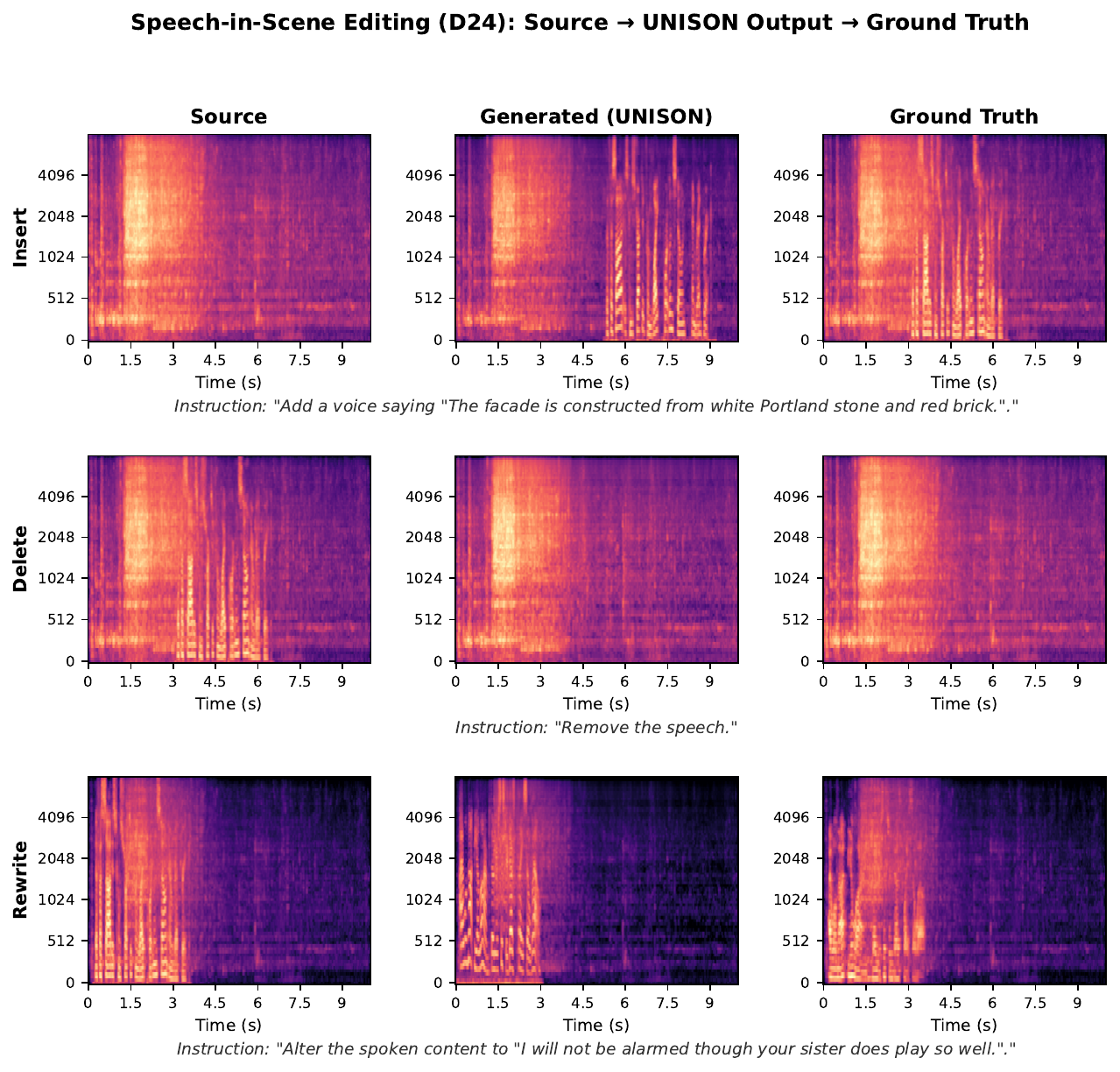}
    \caption{Speech-in-scene editing qualitative examples from \textbf{UNISON (D24, 16\,kHz)}. Each row shows one sub-task (Insert / Delete / Rewrite). The model inserts speech, removes existing speech while preserving the soundscape, or rewrites spoken content.}
    \label{fig:edit_speech_mel_d24}
\end{figure*}

\begin{figure*}[bp]
    \centering
    \includegraphics[width=\linewidth]{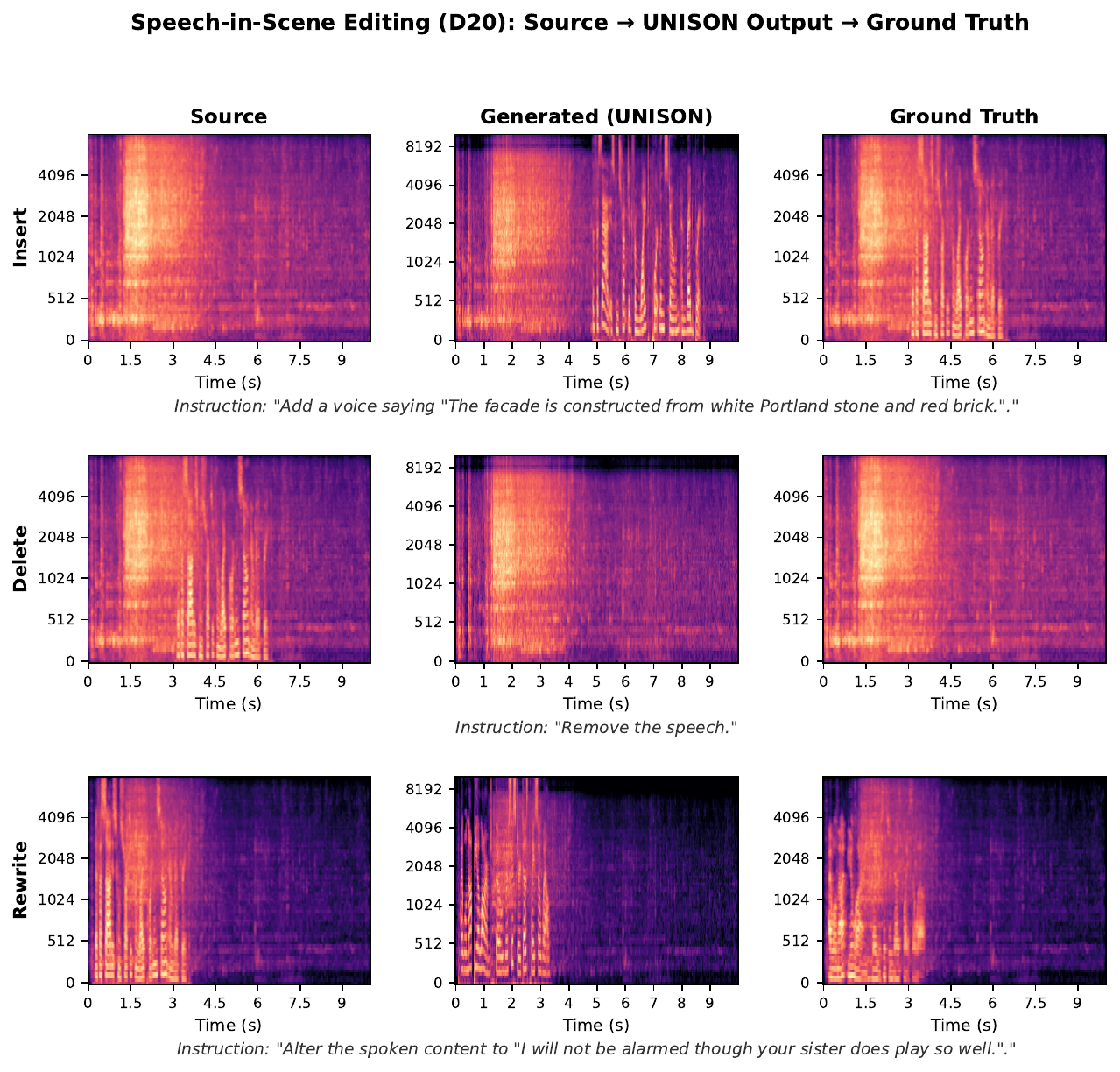}
    \caption{Speech-in-scene editing qualitative examples from \textbf{UNISON (D20, 44.1\,kHz)} on the same samples as Figure~\ref{fig:edit_speech_mel_d24}.}
    \label{fig:edit_speech_mel_d20}
\end{figure*}

Figures~\ref{fig:edit_audio_mel_d24}--\ref{fig:edit_audio_mel_d20} and Figures~\ref{fig:edit_speech_mel_d24}--\ref{fig:edit_speech_mel_d20} present mel spectrogram comparisons for audio editing and speech-in-scene editing tasks from both model variants, respectively. Each figure shows three columns---source audio (input to the model), UNISON's generated output, and the constructed ground truth---for one representative sample per sub-task (add/remove/replace for audio editing; insert/delete/rewrite for speech editing). The instruction text is shown below each row. These visualizations complement the quantitative results in Tables~\ref{tab:audio_edit} and~\ref{tab:speech_edit}, providing intuitive evidence that UNISON preserves non-edited content while accurately executing the specified modification.

\section{Timed Generation Qualitative Examples}
\label{app:timed_qual}

\begin{figure*}[htbp]
    \centering
    \includegraphics[width=\linewidth]{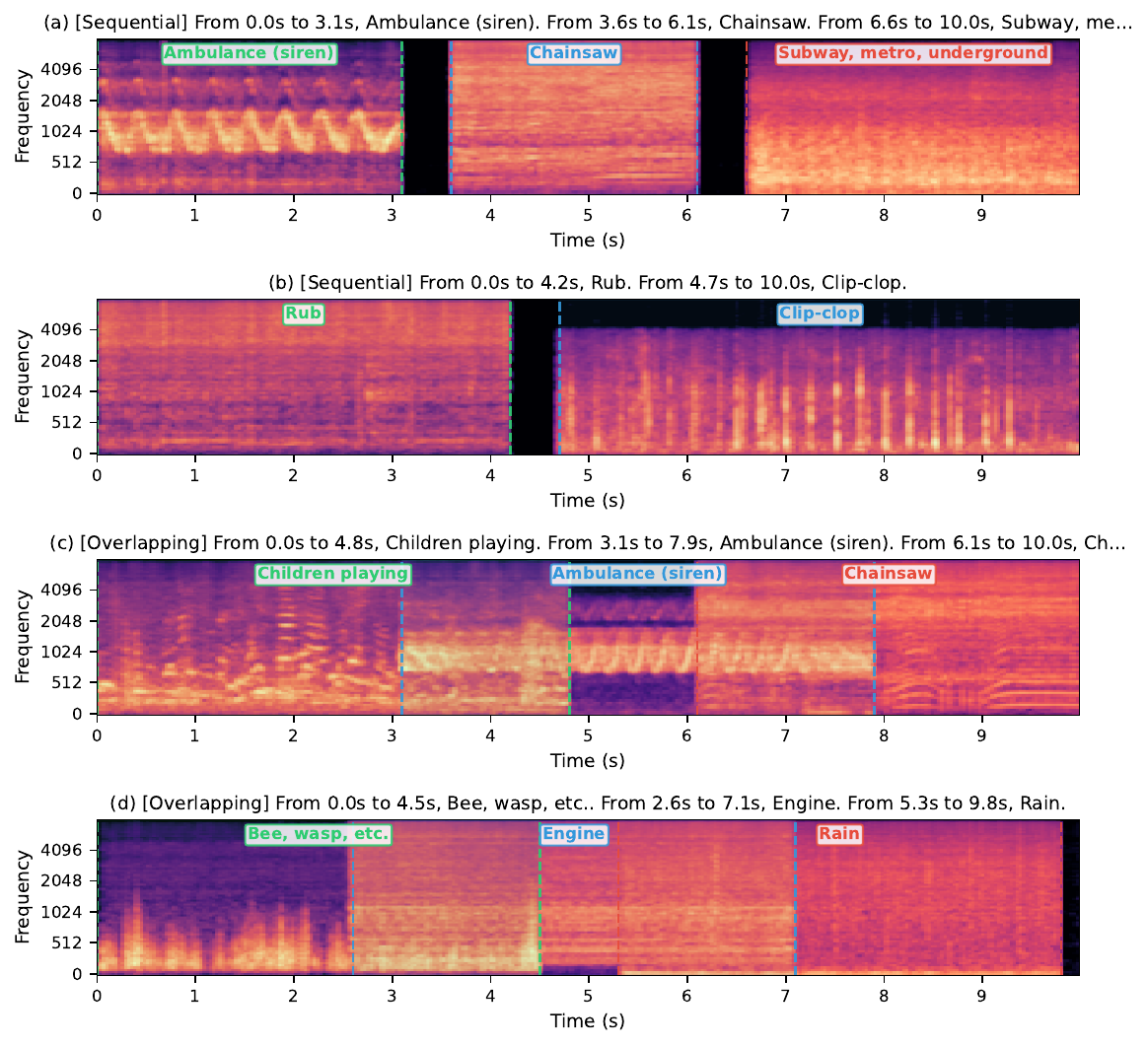}
    \caption{Timed generation mel spectrograms from \textbf{UNISON (D24, 16\,kHz)}. Colored dashed lines and shading denote the time boundaries from the input prompt; segment captions are annotated above each region. (a)--(b): sequential segments. (c)--(d): overlapping segments. The model produces distinct spectral patterns that align with the specified time intervals.}
    \label{fig:timed_mel_d24}
\end{figure*}

\begin{figure*}[htbp]
    \centering
    \includegraphics[width=\linewidth]{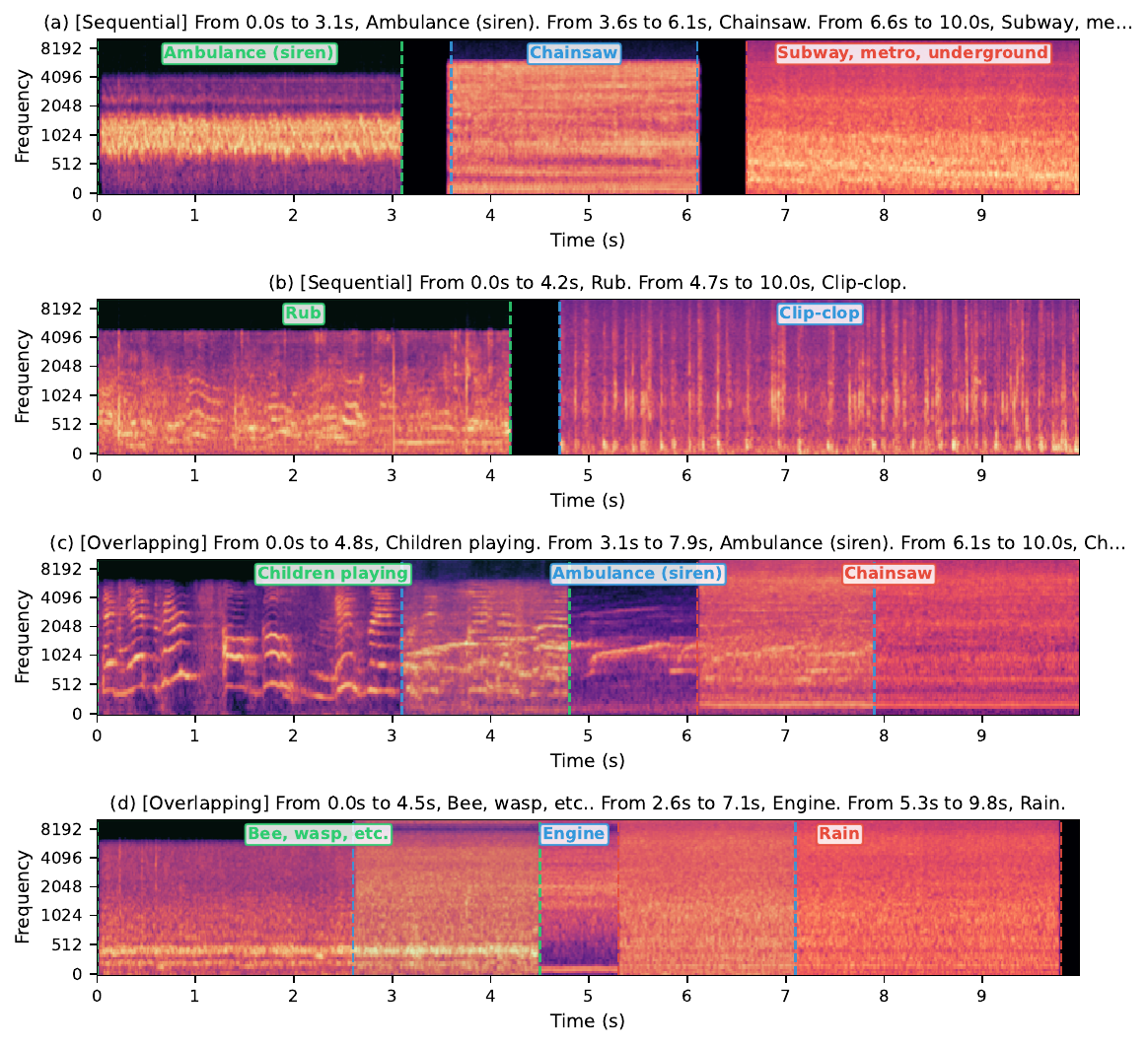}
    \caption{Timed generation mel spectrograms from \textbf{UNISON (D20, 44.1\,kHz)} on the same prompts as Figure~\ref{fig:timed_mel_d24}. The higher sample rate and 40-channel VAE produce richer spectral detail, particularly in the upper frequency bands. Temporal alignment with prompt boundaries remains consistent across both model variants.}
    \label{fig:timed_mel_d20}
\end{figure*}

Figures~\ref{fig:timed_mel_d24} and~\ref{fig:timed_mel_d20} visualize mel spectrograms of timed generation outputs from both model variants. Each panel corresponds to one generated sample; dashed vertical lines and colored shading mark the time boundaries specified in the natural-language prompt, with segment captions annotated above. The top two rows show sequential (non-overlapping) prompts, while the bottom two show overlapping segments. Across both models, the spectrograms confirm that distinct spectral patterns activate within the specified time intervals---validating that UNISON's temporal control operates purely through frozen-LLM instruction parsing without explicit alignment modules.

\end{document}